\documentclass[a4paper,pra,twocolumn]{revtex4}

\usepackage{graphicx}

\begin{document}
%
%
\title{Default-off inter-qubit interactions in NMR quantum computing in rotating solids}
%
%
\author{Kazuyuki Takeda}
\author{Hiromitsu Tanabe}
\author{Masahiro Kitagawa}
\affiliation{Division of Advanced Electronics and Optical Science, Graduate School of Engineering Science, Osaka University, Toyonaka, Osaka 560-8531, Japan}
\date{\today}
\begin{abstract}
A dipolar recoupling technique is introduced as a new approach to quantum gate operation in solid-state NMR under magic angle spinning. The default-off property of inter-qubit interaction provides a simple way to controlled operation without requiring elaborate qubit-decoupling pulses.
\end{abstract}
%
%
%
%
%
%
\maketitle
%
%
\section{Introduction}
In NMR quantum computing experiments demonstrated so far, quantum gate operation has been realized by eliminating all qubit interactions except for the one that is intended to drive a specific quantum gate. The property that the qubit interactions are switched on ``by default'' requires one to apply persistent and sometimes elaborate qubit-decoupling pulse sequences in order to compensate the time evolution of the system due to the unwanted interactions. Despite that an architecture of decoupling-free NMR quantum computing has been proposed\cite{goto2003}, it has not yet been realized. In this work we put forth the first experimental demonstration of two-qubit gate operation driven by the ``default-off'' inter-qubit interactions in solid-state NMR.

In the present approach, dipolar interactions among the spins are eliminated by {\em magic angle spinning} (MAS)\cite{andrew1958,lowe1959}, which is widely used to realize high-resolution NMR in solids.
Then, in order to drive a two-qubit gate, the dipolar interaction between an arbitrary, specific spin pair is {\em selectively} recovered using a homonuclear dipolar recoupling technique called {\it Rotational Resonance in the Tilted Rotating frame} (R2TR)\cite{takegoshi1995,takegoshi1997}.
We demonstrate two-qubit gate operation using the R2TR technique in a single crystal sample of $^{13}$C-labeled glycine, in which two $^{13}$C nuclei serve as qubits, while the protons are continuously decoupled by intense RF irradiation.

Since R2TR employs only RF irradiation under stable spinning of the sample, abrupt switching of the dipolar interaction between a specific spin pair is possible, while all the other spins are left decoupled by MAS.
The absence of the irrelevant interactions makes pulse sequences for quantum circuits natural and simple, since there is no need to apply the continuous qubit-decoupling pulses which is otherwise required to compensate the unitary evolution due to these irrelevant couplings\cite{jones1999b}.


\section{Principle}
Under MAS, a dipolar interaction ${\cal H}_{\rm D}$ between homonuclear spins $I$ and $S$ ($I=S=\frac{1}{2}$) is represented as\cite{maricq1979,mehring}
\begin{equation}
  {\cal H}_{\rm D} = D(t) \left[ I_z S_z - \frac{1}{4}(I_+ S_- + I_- S_+) \right].
\end{equation}
Here, $D(t)$ is the time-dependent component due to MAS, and is written as
\begin{equation}
 D(t) = \omega_{{\rm d}1} \cos[\omega_{\rm R} t + \phi] +
        \omega_{{\rm d}2} \cos[2(\omega_{\rm R} t + \phi)],
\end{equation}
where $\omega_{\rm R}$ is a sample spinning frequency, $\phi$ is the initial phase of the spinning sample container, and
\begin{eqnarray}
 \omega_{{\rm d}1} &=& \sqrt{2} \frac{\mu_0}{4\pi} \frac{\gamma^2 \hbar}{r^3}  \sin 2 \theta_{\rm D}, \\
 \omega_{{\rm d}2} &=& \frac{\mu_0}{4\pi} \frac{\gamma^2 \hbar}{r^3} \sin^2 \theta_{\rm D}.
\end{eqnarray}
$\gamma$ is the gyromagnetic ratio, and $r$ is the internuclear distance between $I$ and $S$. $\theta_{\rm D}$ is the angle between the internuclear vector and the sample-spinning axis.
${\cal H}_{\rm D}$ is thus modulated by MAS, and, according to the average Hamiltonian theory\cite{haeberlen1968}, the dipolar interaction is averaged out and does not affect the time evolution of the system in first order when the spinning frequency $\omega_{\rm R}$ exceeds the magnitude of the dipolar coupling.

Under RF irradiation with an intensity $\omega_1$ at a frequency $\omega$, the Hamiltonian of the spin system in a reference frame rotating around the static magnetic field at $\omega$ is represented as
\begin{equation}
  {\cal H} = \Delta \omega_I I_z + \omega_1 I_x + \Delta \omega_S S_z + \omega_1 S_x + {\cal H}_{\rm D}.
\end{equation}
Here, $\Delta \omega_I = \omega_{0I}-\omega$ and $\Delta \omega_S = \omega_{0S}-\omega$ are the resonance offsets for the $I$ and $S$ spins, where $\omega_{0I}$ and $\omega_{0S}$ are the isotropic resonance frequencies.
In the rotating frame, $I$ and $S$ feel the effective fields $\omega_{{\rm e}I} = [\Delta \omega_I^2 +\omega_1^2]^{1/2}$ and $\omega_{{\rm e}S} = [\Delta \omega_S^2 +\omega_1^2]^{1/2}$, which make angles $\beta_I = \tan^{-1}(\Delta\omega_I / \omega_1)$ and $\beta_S = \tan^{-1}(\Delta\omega_S / \omega_1)$ with respect to the static field, respectively.

\begin{figure}[tb]
\begin{center}
  \includegraphics{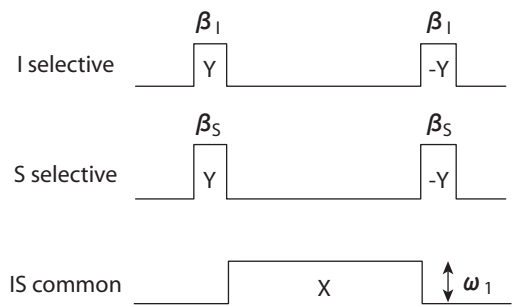}
  \caption{\label{fig1}A pulse sequence for the R2TR experiment.}
\end{center}
\end{figure}

R2TR realizes dipolar recoupling in the tilted rotating frame in which the effective fields for both spins align along the z direction.
The transformation into the tilted rotating frame is realized by a unitary operator
\begin{equation}
  U = \exp(-i \beta_I I_y) \exp(-i \beta_S S_y).
\end{equation}
Experimentally, this is accomplished by applying trim pulses of tip angles $\beta_\xi$ and $- \beta_\xi$ $(\xi=I,S)$ before and after the RF irradiation, respectively, as described in Fig.~\ref{fig1}.

%
%
%
%
%

Among several classes of the recoupling conditions of R2TR, only the followings are of interest, which do not re-introduce the unwanted chemical shift anisotropies.
\begin{eqnarray}
(a) & \omega_{{\rm e}I} - \omega_{{\rm e}S} &= \pm m \omega_{\rm R} \quad (m=1,2), \\
(b) & \omega_{{\rm e}I} + \omega_{{\rm e}S} &= \pm m \omega_{\rm R}.
\end{eqnarray}
The zeroth-order average Hamiltonian for each of these recoupling conditions is
\begin{eqnarray}
(a) & \frac{1}{2} B \omega_{dm} [ I_+ S_- \exp(-i m \phi)  + I_- S_+ \exp(+i m \phi)], \label{eq:b} \\
(b) & \frac{1}{2} Q \omega_{dm} [ I_+ S_+ \exp(-i m \phi) + I_- S_- \exp(+i m \phi)], \label{eq:q} \\
\end{eqnarray}
with
\begin{eqnarray}
  B &=& -\frac{1}{8}(1 + \cos\beta_I \cos\beta_S - 2 \sin\beta_I \sin\beta_S), \\
  Q &=& \frac{1}{8}(1 - \cos\beta_I \cos\beta_S + 2 \sin\beta_I \sin\beta_S).
\end{eqnarray}

For given isotropic resonance frequencies $\omega_{0I}$ and $\omega_{0S}$ and spinning frequency $\omega_{\rm R}$, there are infinite combinations of RF frequency $\omega$ and intensity $\omega_1$ which fulfil either of these recoupling conditions.
It may be helpful to follow the selection tips for efficient recoupling proposed by Takegoshi et al., which are summarized as follows\cite{takegoshi1997}: \\
(i) Recoupling by the ``flip-flop'' mechanism (Eq.~(\ref{eq:b})) is suitable for spins with a chemical shift difference $|\omega_{0I} - \omega_{0S}|$ larger than the dipolar coupling constant $\omega_{\rm D}$, while that by the ``flop-flop'' mechanism (Eq.~(\ref{eq:q})) is desirable for $|\omega_{0I} - \omega_{0S}|$ comparable to or smaller than $\omega_{\rm D}$. \\
(ii) The off-resonance frequencies $\Delta\omega_{I}$ and $\Delta\omega_{S}$ should be much higher than the RF intensity $\omega_1$ in the flip-flop condition, while they should be much lower than $\omega_1$ for the flop-flop condition.

In order to perform a universal quantum gate operation using a two-qubit interaction, the interaction has to be capable of converting a direct-product state into an entangled state\cite{bremner2002}. It has been shown that any arbitrary unitary transformations which create entanglement are expressed in the following form\cite{bremner2002}:
\begin{eqnarray}
  U &=& (A_1 \otimes B_1) \nonumber \\
    & & \times \exp[i (\theta_x I_x \otimes S_x + \theta_y I_y \otimes S_y + \theta_z I_z \otimes S_z)] \nonumber \\
    & & \times (A_2 \otimes B_2). \label{eq:u}
\end{eqnarray}
Here, $A_j, B_j (j=1,2)$ are one-qubit gates for the $I$ and $S$ spins, and at least one of $\theta_x, \theta_y, \theta_z$ must not be zero. It is neither allowed that $\theta_x = \theta_y = \theta_z = \frac{\pi}{4}$. From Eq.~(\ref{eq:u}), the general expression for the Hamiltonian that can drive a universal gate is represented as
\begin{equation}
  \theta_x I_x \otimes S_x + \theta_y I_y \otimes S_y + \theta_z I_z \otimes S_z. \label{eq:uniH}
\end{equation}
Since the recoupled dipolar interaction by R2TR given in Eqs.~(\ref{eq:b}) and (\ref{eq:q}) fits into this group of universal Hamiltonian, the present approach is capable of implementing any quantum circuits.

\section{Experimental}
Fully $^{13}$C labeled glycine was dissolved in a distiled water together with normal, unlabeled glycine with a ratio of 1:9, from which single crystals of glycine was obtained by recrystallization. A single crystal with a size of 2.5$\times$2.5$\times$4 mm was put into a 4 mm$\phi$ zirconia rotor with an arbitrary crystal orientation. In order to take mechanical balance, KBr powder was also packed in the rotor. KBr also served as a sample for adjusting the magic angle through observation of the $^{79}$Br NMR signals.

\begin{figure}[tb]
\begin{center}
  \includegraphics[scale=0.7]{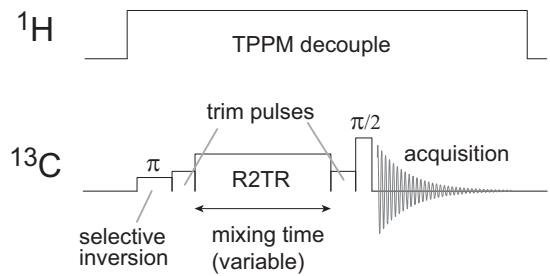}
  \caption{\label{fig2}The pulse sequence for the flip-flop exchange of $^{13}$C magnetizations by R2TR. Throughout the sequence, the unwanted $^{13}$C-$^1$H dipolar interactions are eliminated by TPPM decoupling.}
\end{center}
\end{figure}

Fig.~\ref{fig2} shows a pulse sequence for the R2TR experiment. Firstly, an initial state is prepared by selectively rotating magnetizations of the individual $^{13}$C spins. Secondly, the $^{13}$C spin system is let evolve under the recoupled dipolar interaction by R2TR.
And finally, the resultant $^{13}$C magnetizations are monitored by applying a hard $\pi/2$ pulse before acquiring the FID.
Throughout the sequence, the dipolar interactions with protons, which cannot be removed by MAS with available spinning frequencies, was eliminated with the $^1$H TPPM decoupling technique\cite{bennett1995,gan1997}.

%
%
%
%
%

\section{Result and Discussion}
In order to demonstrate dipolar recoupling of the flip-flop component, either of the two $^{13}$C spins was inverted with a selective $\pi$ pulse before applying the R2TR irradiation with the $m=2$ condition. Fig.~3(a) shows that the $^{13}$C spins exchange their magnetizations with each other under R2TR, confirming that the flip-flop term of the dipolar interaction has been re-introduced under MAS. The period (3.3 msec) of the oscillation indicates that the angle between the internuclear vector and the rotor axis was 64$^\circ$. For comparison, we carried out the same experiment except that the R2TR condition was not satisfied, which resulted in no exchange of the $^{13}$C magnetizations as is demonstrated in Fig.~3(b).

%
%
\begin{figure}[tb]
\begin{center}
  \includegraphics[scale=0.5]{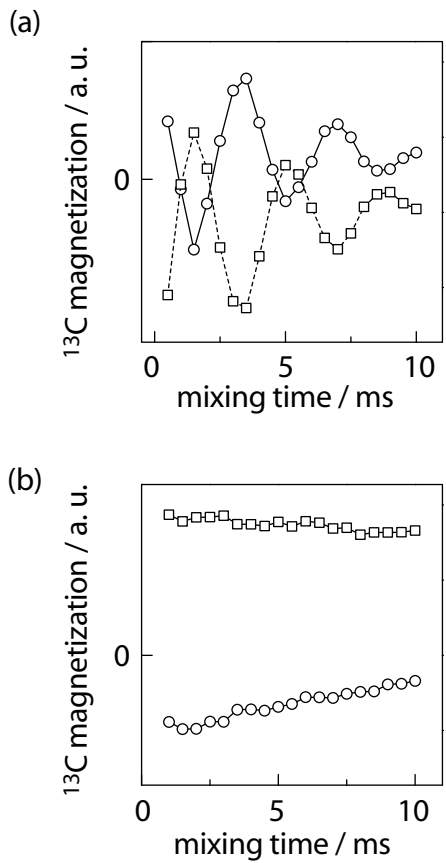}
  \caption{\label{fig3}(a) Exchange of $^{13}$C magnetizations in single crystal of $^{13}$C-labeled glycine in the R2TR experiment. Circles and squares indicate the magnetizations of methylene ($I$) and the carboxyl ($S$) carbon spins, respectively. The selective $\pi$ pulse was initially applied at the methylene carbon spin, and R2TR was then applied with $\omega_1/2\pi=$2339 Hz, $\Delta\omega_I/2\pi=$2000 Hz, $\Delta\omega_S/2\pi=$18699 Hz, and $\omega_{\rm R}/2\pi=$7884 Hz. The R2TR irradiation was sandwiched by the trim pulse of a tip angle $\beta_I$ of 49$^\circ$ for the methylene carbon, while that for the carboxyl carbon was omitted because the required tip angle $\beta_S = 7 ^\circ$ was very small. For comparison, a result with an ``off-R2TR'' condition ($\omega_1/2\pi=$8823 Hz, $\Delta\omega_I/2\pi=$2000 Hz, $\Delta\omega_S/2\pi=$18699 Hz, and $\omega_{\rm R}/2\pi=$7884 Hz) is shown in (b).}
\end{center}
\end{figure}
%
%

Since the interactions between qubits are absent by default, the individual resonance peaks in the readout NMR spectrum do not have multiplet structure under MAS, from which one would be able to tell, in the conventional NMR readout spectrum, the quantum state of the system. This, however, does not mean that the quantum state cannot be extracted in the present case. As described in Fig.~4, the entire spectrum reflects the current quantum state as well. For example, the quantum states $|0\rangle |0\rangle$, $|0\rangle |1\rangle$, $|1\rangle |0\rangle$, and $|1\rangle |1\rangle$ correspond to spectra in Fig.~4(a)-(d), respectively. Extension of this readout scheme to many-qubit systems is straightforward.

Fig.~4(e)-(h) show spectra obtained after half an oscillation period (1.6 ms) of the exchange by R2TR for the initial states of Fig.~4(a)-(d), respectively. As expected, the methylene and the carboxyl carbons exchange their spin states for the initial states $|0\rangle |1\rangle$ and $|1\rangle |0\rangle$, whereas no change was found for the initial states $|0\rangle |0\rangle$ and $|1\rangle |1\rangle$.

%
%
%
\begin{figure}[tb]
\begin{center}
  \includegraphics[scale=0.5]{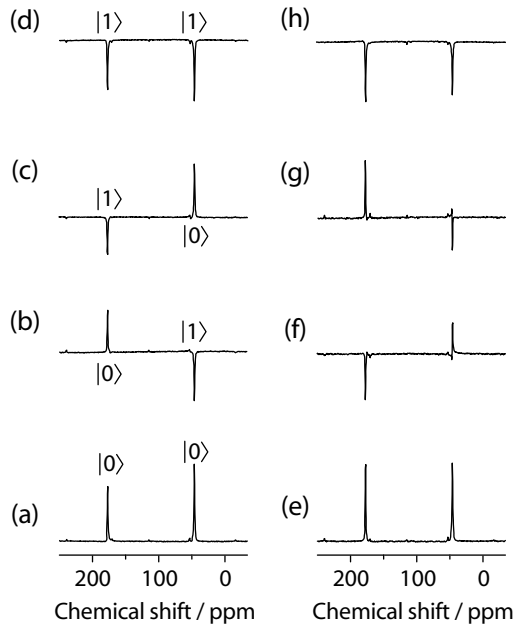}
  \caption{\label{fig4} $^{13}$C NMR spectra in single crystal of $^{13}$C-labeled glycine obtained under MAS and TPPM $^1$H decoupling. (a)-(d) Readout spectra for the quantum states $|0\rangle|0\rangle$, $|0\rangle|1\rangle$, $|1\rangle|0\rangle$, and $|1\rangle|1\rangle$. The left peak (at ca.~177 ppm) and right peak (at ca.~46 ppm) correspond to signals coming from the carboxyl and the methylene carbons. These four states were prepared from the thermal-equilibrium state ($|0\rangle|0\rangle$) by applying a $\pi$ pulse at neither, either, or both spins. (e)-(h) Readout spectra after letting the system evolve under R2TR for half the oscillation period (1.6 ms) from the initial states (a)-(d), respectively. The experiments were carried out in a magnetic field of 11.7 T, and the carrier frequency of $^{13}$C NMR was 125.68 MHz.}

\end{center}
\end{figure}
%
%
%

When the coupling between qubits is given by a Hamiltonian of the form $A(I_+ S_- + I_- S_+)$, the natural choice for the elementary two-qubit gate is the ISWAP gate\cite{schuch2003}, with which quantum circuits can be implemented efficiently. The ISWAP gate, whose matrix representation is given by
\begin{equation}
  \left(
    \begin{array}{cccc}
      1 & 0 & 0 & 0 \\
      0 & 0 & i & 0 \\
      0 & i & 0 & 0 \\
      0 & 0 & 0 & 1 
    \end{array}
  \right)
  =
  \left(
    \begin{array}{cccc}
      1 & 0 & 0 & 0 \\
      0 & 0 & 1 & 0 \\
      0 & 1 & 0 & 0 \\
      0 & 0 & 0 & 1 
    \end{array}
  \right)
  \cdot
  \left(
    \begin{array}{cccc}
      1 & 0 & 0 & 0 \\
      0 & i & 0 & 0 \\
      0 & 0 & i & 0 \\
      0 & 0 & 0 & 1 
    \end{array}
  \right),\label{eq:iswap}
\end{equation}
is realized by letting the system evolve under this Hamiltonian for a period of $\pi/(2A)$. And this ISWAP gate, in turn, can be used to implement the CNS (CNOT + SWAP) gate, which requires only a single operation using this Hamiltonian. In this sense the CNS gate, instead of the CNOT gate, can be regarded as a natural choice for such a case\cite{schuch2003}.

\begin{figure}[tb]
\begin{center}
  \includegraphics[scale=0.5]{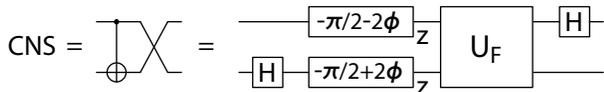}
  \caption{\label{fig5} A quantum circuit for implementing the CNS (CNOT + SWAP) gate with time evolution by the flip-flop Hamiltonian. The matrix representation of $U_{\rm F}$ is given in Eq.~(\ref{eq:phiswap}).}
\end{center}
\end{figure}

This is a special case ($\phi=0$) of the flip-flop Hamiltonian (Eq.~(\ref{eq:b})) in the present study. For an arbitrary initial phase $\phi$ of the rotor, the CNS gate can also be implemented in the following way. Time evolution under Eq.~(\ref{eq:b}) for half the oscillation period $(\pi/\omega_{d2})$ is depicted by a unitary matrix
\begin{equation}
  U_{\rm F} \equiv
  \left(
    \begin{array}{cccc}
      1 & 0 & 0 & 0 \\
      0 & 0 & e^{i(\frac{\pi}{2}-2\phi)} & 0  \\
      0 & e^{i(\frac{\pi}{2}+2\phi)} & 0 & 0 \\
      0 & 0 & 0 & 1 
    \end{array}
  \right), \label{eq:phiswap}
\end{equation}
which, from comparison with Eq.~(\ref{eq:iswap}), is quite similar to ISWAP except for the phase factor $e^{\pm 2 \phi}$ in the non-vanishing off-diagonal elements. It can be shown that this, as well as the ISWAP gate, is equivalent to the CNS gate by calculating the unitary matrix for the quantum circuit given in Fig.~\ref{fig5}.

%
%
%
%
%
%

It has been shown that selective recoupling by R2TR can be efficient when differences in the resonance frequencies of the individual spins exceed ca.~1000 Hz\cite{takegoshi1997}. Thus, it is possible to extend the present approach to systems having more than two homonuclear spins.
There have also been a large number of dipolar recoupling techniques under MAS, some of which are specific to homonuclear recoupling, whereas others are to heteronuclear recoupling.
Among them, only those which introduce the universal Hamiltonian (Eq.~(\ref{eq:uniH})) are of interest in the context of quantum gate operation.
Such techniques include R2TR for homonuclear spins, and rotary resonance recoupling\cite{oas1988} and selective cross polarization\cite{baldus1998} for heteronuclear spins.
We plan to combine these techniques, so that the present approach of default-off controlled operation may be scalable up to several qubits.

\section{Summary}
To summarize, selective dipolar recoupling under MAS was applied to drive two-qubit gates. Since the inter-qubit interactions are eliminated by default, no qubit-decoupling pulses are required. Such a ``default-off'' property of interactions between qubits makes implementation of quantum circuits simple. The present approach is also attractive in the sense that experiments are performed in the solid-state, because there have been solid-state NMR studies in which nuclear spin polarization has been significantly enhanced\cite{takeda2004a,takeda2004b,iinuma2000} to the extent that, when combined with the present approach, one may be able to exploit entanglement.

\section{Acknowledgment}
This work has been supported by the CREST program of Japan Science and Technology Agency.

\bibliography{main.bib}

\end{document}